\documentclass[11pt]{article}
\usepackage{fullpage}
\usepackage{amsmath}
\usepackage{amssymb}
\usepackage{amsfonts}
\usepackage{epsfig}
\usepackage[all,poly]{xy}
\usepackage{colortbl}

\newif\ifnotesw\noteswtrue
   {\ifnotesw\marginpar[\hfill\(\top\)]{\(\top\)}\fi}%
   {\ifnotesw\marginpar[\hfill\(\bot\)]{\(\bot\)}\fi}

\newcommand{\mnote}[1]%
    {\ifnotesw\marginpar%
        [{\scriptsize\begin{minipage}[t]{\marginparwidth}
        \raggedleft#1%
                        \end{minipage}}]%
        {\scriptsize\begin{minipage}[t]{\marginparwidth}
        \raggedright#1%
                        \end{minipage}}%
    \fi}

\newcommand{\ignore}[1]{}

\newcommand{\etal}{{\it et al.~}}

\newsavebox{\given}
\savebox{\given}[1em]{\rule[-1.5ex]{.2mm}{4ex}}

\newcommand{\bnum}{\begin{equation}}
\newcommand{\enum}{\end{equation}}

\newtheorem{theorem}{Theorem}
\newtheorem{corollary}[theorem]{Corollary}
\newtheorem{lemma}[theorem]{Lemma}
\newtheorem{proposition}[theorem]{Proposition}

\newtheorem{fact}[theorem]{Fact}

\newcommand{\blackslug}{\rule{7pt}{7pt}}

\newcommand{\qed}{\hfill{\setlength{\fboxsep}{0pt}
\framebox[7pt]{\rule{0pt}{7pt}}}}

\renewcommand{\notin}{\ifmmode \not\in \else $\not\in$ \fi}
\newlength{\thislabel}
\newcommand{\labsize}[1]{\settowidth{\thislabel}{#1}}

\newcommand{\prf}{\par\noindent{\sl Proof } \hspace{.01 in}}
\newcommand{\zo}{\{0,1\}}

\newcommand{\lip}{\langle}
\newcommand{\rip}{\rangle}

\newcommand{\ZZ}{\mathbb{Z}}

\newcommand{\QQ}{\mathbb{Q}}

\newcommand{\GG}{\mathcal{G}}
\newcommand{\REG}{\mathcal{G}}

\newcommand{\bra}[1]{\lip #1 |}
\newcommand{\ket}[1]{| #1 \rip}
\newcommand{\braket}[2]{\lip #1 | #2 \rip}

\newcommand{\ts}{t^{\star}}

\DeclareMathOperator{\ODD}{ODD-CIRC}
\DeclareMathOperator{\Spec}{Spec}
\DeclareMathOperator{\Integral}{INT}

\title{Perfect state transfer, graph products and equitable partitions 
}
\author{
Yang Ge\footnote{Department of Mathematics, Harvard University.} 
\and
Benjamin Greenberg\footnote{Department of Mathematics, Grinnell College.} 
\and
Oscar Perez\footnote{Centro de Investigaci\'{o}n en Matem\'{a}ticas, Universidad Aut\'{o}noma del Estado de Hidalgo, 
Hidalgo, Mexico.} 
\and
Christino Tamon\footnote{Department of Computer Science, Clarkson University. Contact author: tino@clarkson.edu}
}
\date{\today}
\begin{document}
\maketitle
\bibliographystyle{plain}

\begin{abstract}
We describe new constructions of graphs which exhibit perfect state transfer on continuous-time quantum walks.
Our constructions are based on variants of the double cones \cite{bcms09,anoprt10,anoprt09}
and the Cartesian graph products (which includes the $n$-cube $Q_{n}$) \cite{cddekl05}.
Some of our results include:
\begin{itemize}
\item 
If $G$ is a graph with perfect state transfer at time $t_{G}$, where $t_{G}\Spec(G) \subseteq \ZZ\pi$, and
	$H$ is a circulant with odd eigenvalues, their weak product $G \times H$ has perfect state transfer. 
	Also, if $H$ is a regular graph with perfect state transfer at time $t_{H}$ and $G$ is a graph 
	where $t_{H}|V_{H}|\Spec(G) \subseteq 2\ZZ\pi$, their lexicographic product $G[H]$ has perfect state transfer. 
	For example, these imply $Q_{2n} \times H$ and $G[Q_{n}]$ have perfect state transfer, whenever
	$H$ is any circulant with odd eigenvalues and $G$ is any integral graph, for integer $n \ge 2$.
	These complement constructions of perfect state transfer graphs based on Cartesian products.

\item 
The double cone $\overline{K}_{2} + G$ on any connected graph $G$, has perfect state transfer if 
	the weights of the cone edges are proportional to the Perron eigenvector of $G$. 
	This generalizes results for double cone on regular graphs studied in \cite{bcms09,anoprt10,anoprt09}. 

\item
For an infinite family $\GG$ of regular graphs, 
	there is a circulant connection so the graph $K_{1}+\GG\circ\GG+K_{1}$ has perfect state transfer.
	In contrast, no perfect state transfer exists if a complete bipartite connection is used 
	(even in the presence of weights) \cite{anoprt09}.
	Moreover, we show that the cylindrical cone $K_{1}+\GG+\overline{K}_{n}+\GG+K_{1}$ has no perfect state transfer,
	for any family $\GG$ of regular graphs.

\end{itemize}
We also describe a generalization of the path collapsing argument \cite{ccdfgs03,cddekl05},
which reduces questions about perfect state transfer to simpler (weighted) multigraphs,
for graphs with equitable distance partitions.
Our proofs exploit elementary spectral properties of the underlying graphs.

\hspace{.01in}
\end{abstract}


\section{Introduction}

Recently, perfect state transfer in continuous-time quantum walks on graphs has received considerable attention.
This is due to its potential applications for the transmission of quantum information over quantum networks.
It was originally introduced by Bose \cite{bose03} in the context of quantum walks on linear spin chains
or paths. Another reason for this strong interest is due to the universal property of quantum walks as a computational model
as outlined by Childs \cite{childs09}.
From a graph-theoretic perspective, the main question is whether there is a spectral characterization of graphs which 
exhibit perfect state transfer. Strong progress along these lines were given on highly structured graphs 
by Bernasconi \etal \cite{bgs08} for hypercubic graphs and 
by Ba\v{s}i\'{c} and Petkovi\'{c} \cite{bp09} for integral circulants. 
Nevertheless, a general characterization remains elusive (see Godsil \cite{godsil08}).

Christandl \etal \cite{cdel04,cddekl05} showed that the $n$-fold Cartesian product of the one-link $P_{2}$ and two-link 
$P_{3}$ graphs admit perfect state transfer. This is simply because $P_{2}$ and $P_{3}$ have end-to-end perfect state transfer
and the Cartesian product operator preserves perfect state transfer.
They also drew a crucial connection between hypercubic networks and weighted paths
using the so-called {\em path-collapsing} argument. This argument was also used by Childs \etal \cite{ccdfgs03} 
in the context of an exponential algoritmic speedup for a black-box graph search problem via continuous-time quantum walks. 
Christandl \etal \cite{cddekl05} proved that, although the $n$-vertex path $P_{n}$, for $n \ge 4$, has no end-to-end
perfect state transfer, a suitably weighted version of $P_{n}$ has perfect state transfer (via a path-collapsing
reduction from the $n$-cube $Q_{n}$). A somewhat critical ingredient of this reduction is that each layer of $Q_{n}$
is an empty graph. We generalize this argument to graphs which have equitable distance partitions 
(see Godsil and Royle \cite{godsil-royle01}).

\begin{figure}[t]
\begin{center}
\epsfig{file=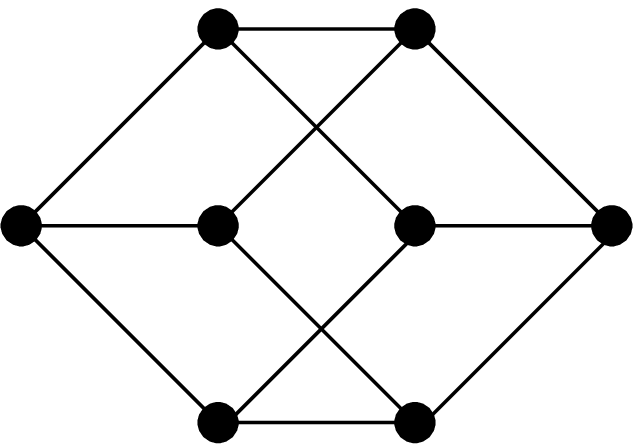, height=1.05in, width=1.65in}
\hspace{.5in}
\epsfig{file=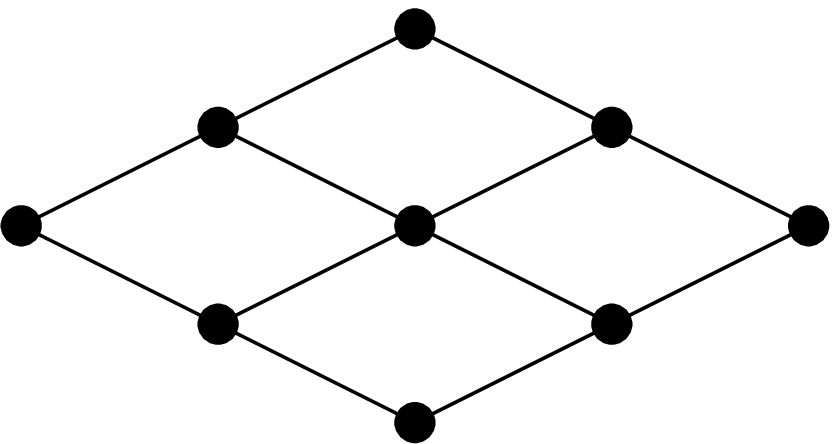, height=1.05in, width=1.95in}
\caption{The Cartesian product construction for perfect state transfer: (a) $P_{2} \oplus P_{2}$; 
(b) $P_{3} \oplus P_{3}$ (see Christandl \etal \cite{cddekl05}).}
\label{fig:cartesian}
\end{center}
\hrule
\end{figure}

Bose \etal \cite{bcms09} observed an interesting phenomena on the complete graph $K_{n}$. Although $K_{n}$ does
not exhibit perfect state transfer, they show that by removing an edge between any two vertices, perfect state transfer 
is created between them. Note that the graph we obtain from removing an edge from $K_{n}$ is the double cone 
$\overline{K}_{2}+K_{n-2}$ (where $G+H$ denotes the join of graphs $G$ and $H$). 
This observation was generalized in Angeles-Canul \etal \cite{anoprt10} where perfect state transfer was proved for 
double cones $\{\overline{K}_{2},K_{2}\}+G$, where $G$ is some regular graph (in place of complete graphs). 
The analyses on these double cones showed that perfect state transfer need not occur between antipodal vertices 
and that having integer eigenvalues is not a sufficient condition for perfect state transfer 
(which answered questions raised in \cite{godsil08}).

Our goal in this work is to combine and extend both the Cartesian product and the double cone constructions. 
The Cartesian product construction (which combines graphs with perfect state transfer) has the advantage of
producing large diameter graphs with antipodal perfect state transfer. In fact, this construction provides the
best upper bound for the order-diameter problem; for a given $d$, let $f(d)$ be the smallest size graph which 
has perfect state transfer between two vertices of distance $d$. Then, the best known bounds are $d \le f(d) \le \alpha^{d}$, 
where $\alpha = 2$, if $d$ is odd, and $\alpha = \sqrt{3}$, if $d$ is even; here, the upper bounds are achieved
by $P_{2}^{\otimes n}$ and $P_{3}^{\otimes n}$. 
On the other hand, the double cone construction allows graphs whose quotients (modulo its equitable partition)
contain cells which are not independent sets. This can potentially allow for a broader class of graphs
with perfect state transfer (see Bose \etal \cite{bcms09} and Angeles-Canul \etal \cite{anoprt10,anoprt09}).

In this work, we describe new constructions of families of graphs with perfect state transfer.
First, we extend several of the double cone constructions and relax their diameter restrictions.
We show that the double cone $\overline{K}_{2}+G$ of an arbitrary connected graph $G$ has perfect state transfer 
if we use edge weights proportional to the Perron eigenvector of $G$. This extends results given in \cite{anoprt10} 
where $G$ is required to be a regular graph. 
Then, we prove that the glued double cone graph $K_{1}+G_{1} \circ G_{2}+K_{1}$ has perfect state transfer
whenever $G_{1},G_{2}$ belongs to some class of regular graphs and if they are connected using some matrix 
$C$ which commutes with the adjacency matrices of $G_{1}$ and $G_{2}$. 
In contrast, Angeles-Canul \etal \cite{anoprt09} proved that $K_{1}+G+G+K_{1}$ has no perfect state transfer, 
for any regular graph $G$, even if weights are allowed. 

\begin{figure}[t]
\begin{center}
\begin{tabular}{|c|c|c|c|} \hline
Graph family					& PST	& Construction    		& Source		\\  \hline
$\{\overline{K}_{2},K_{2}\}+\GG$& yes	& join					& \cite{bcms09,anoprt10} \\
$\{\overline{K}_{2},K_{2}\}+\tilde{\GG}$				
								& yes${}^{\ast}$	& join					& this work \\ \hline
$P_{n \ge 4}$  					& no 	& path					& \cite{cddekl05} \\
$K_{1}+\GG \circ \GG+K_{1}$ 	& yes	& circulant half-join	& this work \\ 
$K_{1}+\GG+\GG+K_{1}$ 			& no${}^{\ast}$	& half-join				& \cite{anoprt09} \\
$K_{1}+\GG+\overline{K}_{n}+\GG+K_{1}$ 	
								&  no  	& join					& this work \\  \hline
$Q_{n}$ or $P_{3}^{\oplus n}$	&  yes	& Cartesian product		& \cite{cddekl05} \\
$\{Q_{2n},P_{3}^{\oplus n}\} \times \ODD$		
								& yes 	& weak product			& this work \\ 
$\Integral[Q_{n \ge 2}]$		& yes	& lexicographic product	& this work \\ \hline
\end{tabular}
\caption{
{\em Summary of results on some graphs with perfect state transfer}:
$n$ is a positive integer;
$\GG$ denotes some family of regular graphs; $\tilde{\GG}$ denotes an arbitrary connected graph;
$P_{n}$ is the path on $n$ vertices; $Q_{n}$ is the $n$-dimensional cube; $K_{n}$ is the complete graph on $n$ vertices;
$\ODD$ is the class of circulant graphs with odd eigenvalues;
$\Integral$ is the class of integral graphs.
Asterisks indicate results on weighted graphs. 
}
\label{fig:results}
\end{center}
\hrule
\end{figure}

For cones with larger diameter, we consider the graph $K_{1}+G_{1}+H+G_{2}+K_{1}$, 
where $G_{1},G_{2}$ belong to the same class of regular graphs and $H$ is another regular graph. 
This symmetry is a necessary condition for perfect state transfer as shown by Kay \cite{kay09}.
Nevertheless, in contrast to the previous positive results, we show there is no perfect state transfer whenever 
$H$ is the empty graph. 
The $4$-dimensional cube $Q_{4}$ (which has perfect state transfer) is an example of such a graph but without 
the join (or complete bipartite) connection.

Our other contribution involves constructions of perfect state transfer graphs using alternative graph products,
namely the weak and lexicographic products. 
An interesting property of these products is that they can create perfect state transfer graphs by combining graphs 
with perfect state transfer and ones which lack the property. For example, we show that $Q_{2n} \times K_{2m}$ has 
perfect state transfer, for any integers $n$ and $m$. Recall that the complete graph has no perfect state transfer 
(as observed by Bose \etal \cite{bcms09}). In comparison, the Cartesian product requires both of its graph arguments 
to have perfect state transfer (with the same perfect state transfer times).
We also consider the lexicographic graph product (or graph composition) and its generalizations. 
Our generalized lexicographic product of $G$ and $H$ using a connection matrix (or graph) $C$ is a graph
whose adjacency matrix is $A_{G} \otimes C + I \otimes A_{H}$. Note we recover the Cartesian product 
by letting $C=I$ and the standard lexicographic product by letting $C=J$.
So, this generalization interpolates between these two known graph products.
For example, we show that $G[Q_{n}]$ has perfect state transfer for any integral graph $G$ and $n \ge 2$.

The proofs we employ exploit elementary spectral properties of the underlying graphs. 
Some of our results are summarized in Figure \ref{fig:results}.


\section{Preliminaries}

Let $[n]$ denote the set $\{0,1,\ldots,n-1\}$.
For a tuple of binary numbers $(a,b) \in \zo^{2} \setminus \{(0,0)\}$, let $\QQ_{a,b}$ denote the set of
rational numbers of the form $p/q$, with $gcd(p,q)=1$, where $p \equiv a\pmod{2}$ and $q \equiv b\pmod{2}$.
These denote rational numbers (in lowest terms) that are ratios of two odd integers or of an odd integer 
and an even integer, or vice versa. We denote the even and odd integers as $2\ZZ$ and $2\ZZ+1$, respectively.

The graphs $G=(V,E)$ we study are finite, simple, undirected, connected, and mostly unweighted.
The adjacency matrix $A_{G}$ of a graph $G$ is defined as $A_{G}[u,v] = 1$ if $(u,v) \in E$ and $0$ otherwise;
we also use $u \sim v$ to mean $u$ is adjacent to $v$.
The spectrum $\Spec(G)$ of $G$ is the set of eigenvalues of $A_{G}$. 
The graph $G$ is called {\em integral} if all of its eigenvalues are integers.
A graph $G=(V,E)$ is called $k$-regular if each vertex $u \in V$ has exactly $k$ adjacent neighbors.
For integers $n \ge 1$ and $0 \le k < n$, let $\REG_{n,k}$ be the set of all $n$-vertex $k$-regular graphs.
The distance $d(a,b)$ between vertices $a$ and $b$ is the length of the shortest path connecting them.

Some standard graphs we consider include the complete graphs $K_{n}$, 
paths $P_{n}$, and circulants graphs. An $n$-vertex {\em circulant} graph $G$ on is a graph whose adjacency 
matrix is an $n \times n$ circulant matrix; that is, there is a sequence $(a_{0},\ldots,a_{n-1})$ so that
$A_{G}[j,k] = a_{k-j}$, where arithmetic on the indices is done modulo $n$. 
Alternatively, we may define a circulant graph $G$ on $[n]$ through a subset $S \subseteq [n]$ 
where $j$ is adjacent to $k$ if and only if $k-j \in S$; we denote such a circulant as $Circ(n,S)$.
Known examples of circulants include the complete graphs $K_{n}$ and cycles $C_{n}$.

Let $G$ and $H$ be two graphs with adjacency matrices $A_{G}$ and $A_{H}$, respectively.
The complement of $G=(V,E)$, denoted $\overline{G}=(V,\overline{E})$, is a graph where $(u,v) \in \overline{E}$ 
if and only if $(u,v) \not\in E$, for $u \neq v$. Some relevant binary graph operations are defined in the following:
\begin{itemize}
\item
The {\em Cartesian product}
$G \oplus H$ is a graph defined on $V(G) \times V(H)$
where $(g_{1},h_{1})$ is adjacent to $(g_{2},h_{2})$ if either
$g_{1}=g_{2}$ and $(h_{1},h_{2}) \in E_{H}$, or $(g_{1},g_{2}) \in E_{G}$ and $h_{1}=h_{2}$.
The adjacency matrix of $G \oplus H$ is $A_{G} \otimes I + I \otimes A_{H}$. 

\item
The {\em weak product} $G \times H$ is a graph defined on $V(G) \times V(H)$
where $(g_{1},h_{1})$ is adjacent to $(g_{2},h_{2})$ if $(g_{1},g_{2}) \in E_{G}$ and $(h_{1},h_{2}) \in E_{H}$.
The adjacency matrix of $G \times H$ is $A_{G} \otimes A_{H}$.

\item
The {\em lexicographic product} $G[H]$ is a graph defined on $V(G) \times V(H)$
where $(g_{1},h_{1})$ is adjacent to $(g_{2},h_{2})$ if either
$(g_{1},g_{2}) \in E_{G}$ or $g_{1}=g_{2}$ and $(h_{1},h_{2}) \in E_{H}$.
The adjacency matrix of $G[H]$ is $A_{G} \otimes J + I \otimes A_{H}$.

\item 
The {\em join} $G + H$ is a graph defined on $V(G) \cup V(H)$ obtained by taking two disjoint copies
of $G$ and $H$ and by connecting all vertices of $G$ to all vertices of $H$.
The adjacency matrix of $G+H$ is $\begin{bmatrix} A_{G} & J \\ J & A_{H} \end{bmatrix}$.

\end{itemize}
We assume appropriate dimensions on the identity $I$ and all-one $J$ matrices used above.
The $n$-dimensional hypercube $Q_{n}$ may be defined recursively as 
$Q_{1} = K_{2}$ and $Q_{n} = K_{2} \oplus Q_{n-1}$, for $n \ge 2$.
The {\em cone} of a graph $G$ is defined as $K_{1} + G$. The {\em double cone} of $G$ is 
$\overline{K}_{2} + G$, whereas the {\em connected} double cone is $K_{2} + G$.

A partition $\pi$ of a graph $G=(V,E)$ given by $V = \biguplus_{j=1}^{m} V_{j}$ is called {\em equitable} if
the number of neighbors in $V_{k}$ of a vertex $u$ in $V_{j}$ is a constant $d_{j,k}$, independent of $u$
(see \cite{godsil-royle01,godsil93}).
The {\em quotient} graph of $G$ over $\pi$, denoted by $G/\pi$, is the directed graph with the $m$ cells of $\pi$ 
as its vertices and $d_{j,k}$ edges from the $j$th to the $k$th cells of $\pi$.
The adjacency matrix of $G/\pi$ is given by $A_{G/\pi}[j,k] = d_{j,k}$.

\begin{figure}[t]
\begin{center}
\epsfig{file=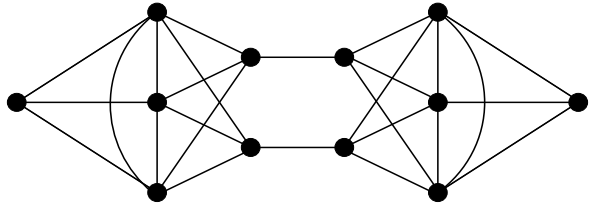, height=1.05in, width=3.05in}
\caption{A cylindrical cone with an equitable distance partition: 
$K_{1}+K_{3}\circ \overline{K}_{2} \circ \overline{K}_{2} \circ K_{3}+K_{1}$.}
\label{fig:cylindrical-cone}
\end{center}
\hrule
\end{figure}

A graph $G$ has an {\em equitable distance partition} $\pi$ with respect to a vertex $a$ if
$\pi = \biguplus_{j=0}^{m} V_{j}$ is such that $G/\pi$ is a {\em path} 
and $V_{j} = \{x \in V : d(x,a) = j\}$ where $V_{0}=\{a\}$;
typically, we also require that there is a vertex $b$, antipodal to $a$, so that $V_{m}=\{b\}$.
We also call a graph a {\em cylindrical cone} (see Figure \ref{fig:cylindrical-cone}) if it has an equitable distance partition
and is denoted $K_{1} \circ G_{1} \circ \ldots \circ G_{m} \circ K_{1}$, where $G_{j}$ are regular graphs 
and $\circ$ denote (semi-)regular bipartite connections (induced by the equitable partition $\pi$).

Further background on algebraic graph theory may be found in the comprehensive texts of Biggs \cite{biggs}, 
Godsil and Royle \cite{godsil-royle01}, and Godsil \cite{godsil93}.

Next, we describe the continuous-time quantum walk as defined originally by Farhi and Gutmann \cite{fg98}.
For a graph $G=(V,E)$, let $\ket{\psi(0)} \in \mathbb{C}^{|V|}$ be an initial amplitude vector of unit length. 
Using Schr\"{o}dinger's equation, the amplitude vector of the quantum walk at time $t$ is
\begin{equation}
\ket{\psi(t)} = e^{-it A_{G}} \ket{\psi(0)}.
\end{equation}
Note since $A_{G}$ is Hermitian (in our case, symmetric), $e^{-itA_{G}}$ is unitary (hence, an isometry).
More detailed discussion of quantum walks on graphs can be found in the excellent surveys by Kempe \cite{kempe03} 
and Kendon \cite{k06}.
The {\em instantaneous} probability of vertex $a$ at time $t$ is $p_{a}(t) = |\braket{a}{\psi(t)}|^{2}$.
We say $G$ has {\em perfect state transfer} from vertex $a$ to vertex $b$ at time $t$ if a continuous-time
quantum walk on $G$ from $a$ to $b$ has unit fidelity or
\begin{equation} \label{eqn:pst}
|\bra{b}e^{-itA_{G}}\ket{a}| = 1,
\end{equation}
where $\ket{a}$, $\ket{b}$ denote the unit vectors corresponding to the vertices $a$ and $b$,
respectively. The graph $G$ has perfect state transfer if there exist vertices $a$ and $b$ in $G$
and time $t$ so that (\ref{eqn:pst}) is true.


\section{Graph products}

In this section, we describe constructions of perfect state transfer graphs using the weak and lexicographic
products. These complement the well-known Cartesian product constructions \cite{cddekl05}.

\subsection{Weak product}

An interesting property of the weak product graph operator is that it can create graphs with perfect state transfer 
by combining ones with perfect state transfer and ones which lack the property. In contrast, the Cartesian graph 
product can only create perfect state transfer graphs from ones which have the property.
We start with the following simple observation.

\begin{fact}
Let $G$ be an $n$-vertex graph and $H$ be an $m$-vertex graph whose eigenvalues and eigenvectors are given by
$A_{G}\ket{u_{k}} = \lambda_{k}\ket{u_{k}}$, for $k \in [n]$, and
$A_{H}\ket{v_{\ell}} = \mu_{\ell}\ket{v_{\ell}}$, for $\ell \in [m]$, respectively.
Let $g_{1},g_{2} \in G$ and $h_{1},h_{2} \in H$.
Then, the fidelity of a quantum walk on their weak product $G \times H$ between $(g_{1},h_{1})$ and $(g_{2},h_{2})$ 
is given by
\begin{eqnarray}
\bra{g_{2},h_{2}}e^{-itA_{G \times H}}\ket{g_{1},h_{1}}
	\label{eqn:weak-fidelity}
	& = & \bra{g_{2}} \left[\sum_{k} 
			\left\{\sum_{\ell} \braket{h_{2}}{v_{\ell}}\braket{v_{\ell}}{h_{1}} e^{-it\lambda_{k}\mu_{\ell}} \right\}
			\ket{u_{k}}\bra{u_{k}} \right] \ket{g_{1}}.
\end{eqnarray}
\end{fact}
\prf
Recall that the adjacency matrix of $A_{G \times H}$ is $A_{G} \otimes A_{H}$. Thus,
the eigenvalues and eigenvectors of the weak product $G \times H$ are
\begin{equation}
A_{G \times H}(\ket{u_{k}} \otimes \ket{v_{\ell}}) = \lambda_{k}\mu_{\ell}(\ket{u_{k}} \otimes \ket{v_{\ell}}),
\ \ \mbox{ where $k \in [n]$ and $\ell \in [m]$. }
\end{equation}
So, the quantum walk on $G \times H$ from $(g_{1},h_{1})$ to $(g_{2},h_{2})$ is given by
\begin{eqnarray}
\bra{g_{2},h_{2}}e^{-itA_{G \times H}}\ket{g_{1},h_{1}}
	& = & \sum_{k,\ell} \braket{g_{2}}{u_{k}}\braket{u_{k}}{g_{1}} \braket{h_{2}}{v_{\ell}}\braket{v_{\ell}}{h_{1}}
			e^{-it\lambda_{k}\mu_{\ell}}.
\end{eqnarray}
After rearranging summations, we obtain the claim.
\qed

\begin{figure}[t]
\begin{center}
\epsfig{file=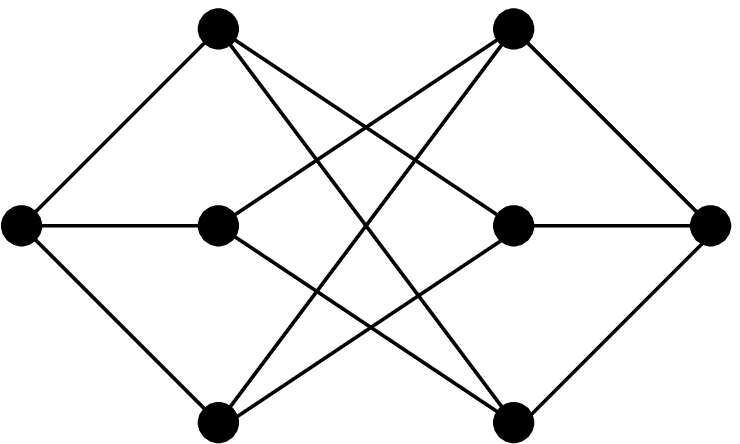, height=1.15in, width=1.75in}
\hspace{.5in}
\epsfig{file=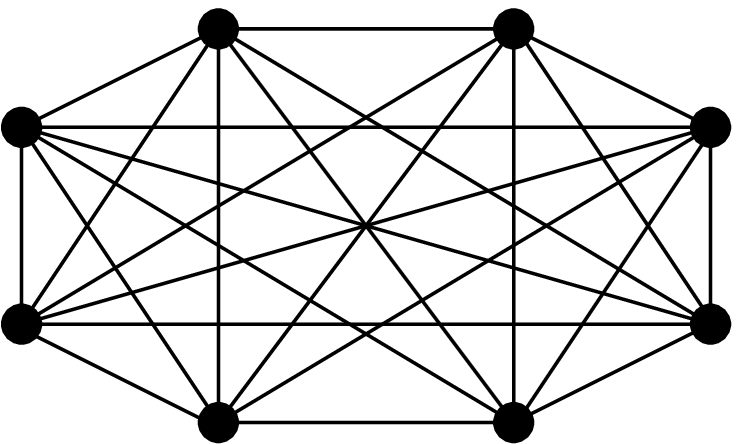, height=1.15in, width=1.75in}
\caption{Graph products with perfect state transfer:
(a) the weak product $K_{2} \times K_{4m}$, for $m \ge 1$ (shown here with $m=1$);
(b) the lexicographic product (or composition) $K_{m}[Q_{n}]$
(shown here with $m=n=2$).
}
\label{fig:graph-product}
\end{center}
\hrule
\end{figure}

\begin{proposition} \label{prop:weak-pst}
Let $G$ be a graph with perfect state transfer at time $t_{G}$ so that
\begin{equation}
t_{G}\Spec(G) \subseteq \ZZ\pi.
\end{equation}
Then, $G \times H$ has perfect state transfer if $H$ is a circulant graph with odd eigenvalues.
\end{proposition}
\prf
Suppose $G$ is an $n$-vertex graph whose eigenvalues and eigenvectors are given by
$A_{G}\ket{u_{k}} = \lambda_{k}\ket{u_{k}}$, for $k \in [n]$.
Assume that $G$ has perfect state transfer at time $t_{G}$ from vertex $g_{1}$ to $g_{2}$.
Also, suppose $H$ be an $m$-vertex graph whose eigenvalues and eigenvevtors are given by
$A_{H}\ket{v_{\ell}} = \mu_{\ell}\ket{v_{\ell}}$, for $\ell \in [m]$.
In Equation (\ref{eqn:weak-fidelity}), if $H$ is circulant on $m$ vertices, we have
$\braket{0}{v_{\ell}}\braket{v_{\ell}}{0} = 1/m$.
Moreover, if each eigenvalue of $H$ is odd, say $\mu_{\ell}=2m_{\ell}+1$, with $m_{\ell} \in \ZZ$, then
\begin{eqnarray}
\bra{g_{2},0} \exp(-it_{G}A_{G \times H}) \ket{g_{1},0}
	& = & \frac{1}{m} \sum_{k,\ell} \braket{g_{2}}{u_{k}}\braket{u_{k}}{g_{1}} e^{-it_{G}\lambda_{k}\mu_{\ell}} \\
	& = & \frac{1}{m} \sum_{k} \braket{g_{2}}{u_{k}}\braket{u_{k}}{g_{1}} 
		\sum_{\ell} e^{-it_{G}\lambda_{k}(2m_{\ell}+1)} \\
	& = & \sum_{k} \braket{g_{2}}{u_{k}}\braket{u_{k}}{g_{1}} 
		e^{-it_{G}\lambda_{k}}, \ \ \mbox{ since $t_{G}\lambda_{k} \in \ZZ\pi$ }.
\end{eqnarray}
The last expression equals to $\bra{g_{2}}e^{-it_{G}A_{G}}\ket{g_{1}}$, by the spectral theorem.
This proves the claim.
\qed\\

\par\noindent{\em Remark}:
Note $Q_{2n}$ has eigenvalues $\lambda_{k} = 2n-2k$, for $k=0,\ldots,2n$, and perfect state transfer time $t=\pi/2$.
Also, $P_{3}^{\otimes n}$ has eigenvalues from $\lambda_{k} \in \ZZ\sqrt{2}$ and perfect state transfer time $t=\pi/\sqrt{2}$. 
In both cases, we have $t\lambda_{k} \in \ZZ\pi$, for all $k$. Thus, by Proposition \ref{prop:weak-pst}, we get that
$\{Q_{2n},P_{3}^{\otimes n}\} \times H$ has perfect state transfer for any circulant $H$ with odd eigenvalues.
For example, we may let $H = K_{m}$ be the complete graph of order $m$, for an even integer $m$.


\subsection{Lexicographic products}

The generalized lexicographic product $G_{C}[H]$ between a graph $G$ and two graphs $H$ and $C$,
with $V_{H} = V_{C}$, is a graph on $V_{G} \times V_{H}$ where
$(g_{1},h_{1})$ is adjacent to $(g_{2},h_{2})$ if and only if either
$(g_{1},g_{2}) \in E_{G}$ and $(h_{1},h_{2}) \in E_{C}$, or,
$g_{1} = g_{2}$ and $(h_{1},h_{2}) \in E_{H}$.
In terms of adjacency matrices, we have
\begin{equation} \label{eqn:generalized-lexico}
A_{G_{C}[H]} = A_{G} \otimes A_{C} + I \otimes A_{H}.
\end{equation}

\par\noindent
We describe constructions of perfect state transfer graphs using generalized lexicographic products.
Again, we start with the following simple observation.

\begin{fact} \label{fact:generalized-lexico}
Let $G$ be an $n$-vertex graph 
whose eigenvalues and eigenvectors are given by
$A_{G}\ket{u_{k}} = \lambda_{k}\ket{u_{k}}$, for $k \in [n]$.
Let $H$ and $C$ be $m$-vertex graphs whose adjacency matrices commute, that is $[A_{H},A_{C}] = 0$,
and whose eigenvalues and eigenvectors are given by
$A_{H}\ket{v_{\ell}} = \mu_{\ell}\ket{v_{\ell}}$, and $A_{C}\ket{v_{\ell}} = \gamma_{\ell}\ket{v_{\ell}}$, 
for $\ell \in [m]$, respectively. 
Suppose $g_{1},g_{2} \in G$ and $h_{1},h_{2} \in H$.
Then, the fidelity of a quantum walk on the generalized lexicographic product $G_{C}[H]$ 
between $(g_{1},h_{1})$ and $(g_{2},h_{2})$ is given by
\begin{equation} \label{eqn:pst-lexico}
\bra{g_{2},h_{2}}\exp(-itA_{G_{C}[H]})\ket{g_{1},h_{1}}
	= \sum_{k} \braket{g_{2}}{u_{k}}\braket{u_{k}}{g_{1}} 
			\sum_{\ell} \braket{h_{2}}{v_{\ell}}\braket{v_{\ell}}{h_{1}}
			e^{-it(\lambda_{k}\gamma_{\ell}+\mu_{\ell})} 
\end{equation}
\end{fact}
\prf
The eigenvalues and eigenvectors of $G_{C}[H]$ are given by
\begin{equation}
A_{G_{C}[H]}(\ket{u_{k}} \otimes \ket{v_{\ell}}) = (\lambda_{k}\gamma_{\ell}+\mu_{\ell})(\ket{u_{k}} \otimes \ket{v_{\ell}}),
\ \ \mbox{ $k \in [n]$ and $\ell \in [m]$}.
\end{equation}
So, the quantum walk on $G_{C}[H]$ from $(g_{1},h_{1})$ to $(g_{2},h_{2})$ is given by
\begin{eqnarray} 
\bra{g_{2},h_{2}}e^{-itA_{G_{C}[H]}}\ket{g_{1},h_{1}}
	& = & \sum_{k,\ell} \braket{g_{2}}{u_{k}}\braket{u_{k}}{g_{1}} 
			\braket{h_{2}}{v_{\ell}}\braket{v_{\ell}}{h_{1}}
			e^{-it(\lambda_{k}\gamma_{\ell}+\mu_{\ell})},
\end{eqnarray}
which proves the claim.
\qed\\

\par\noindent
In the following, we show a closure property of perfect state transfer graphs using a
generalized lexicographic product with the complete graph as a connection matrix.
This is similar to the weak product construction from Proposition \ref{prop:weak-pst}.

\begin{proposition} \label{prop:lexico-clique}
Let $G$ and $H$ be perfect state transfer graphs with a common time $t$.
Assume $H$ is a $m$-vertex graph which commutes with $K_{m}$. 
Suppose that
\begin{equation}
t|V_{H}|\Spec(G) \subseteq 2\ZZ\pi.
\end{equation}
Then, the lexicographic product $G_{K_{m}}[H]$ has perfect state transfer at time $t$.
\end{proposition}
\prf
Suppose $G$ has perfect state transfer from $g_{1}$ to $g_{2}$ at time $t$, where $g_{1},g_{2} \in V_{G}$. 
Let the eigenvalues and eigenvectors of $G$ be given by $A_{G}\ket{u_{k}} = \lambda_{k}\ket{u_{k}}$, for $k \in [n]$.
Also, suppose $H$ is a circulant with perfect state transfer from $h_{1}$ to $h_{2}$ at time $t$, 
where $h_{1},h_{2} \in V_{H}$. Let the eigenvalues and eigenvectors of $H$ be given by
$A_{H}\ket{v_{\ell}} = \mu_{\ell}\ket{\mu_{\ell}}$, for $\ell \in [m]$.
Thus, Equation (\ref{eqn:pst-lexico}) becomes
\begin{eqnarray}
\lefteqn{\bra{g_{2},h_{2}}e^{-itA_{G_{K_{m}}[H]}}\ket{g_{1},h_{1}}}\\
	& = & \sum_{k} \braket{g_{2}}{u_{k}}\braket{u_{k}}{g_{1}}
			\left\{ 
				e^{-it(\lambda_{k}(m-1)+\mu_{0})} \braket{h_{2}}{v_{0}}\braket{v_{0}}{h_{1}} 
				+
				\sum_{\ell \neq 0} e^{-it(-\lambda_{k}+\mu_{\ell})} \braket{h_{2}}{v_{\ell}}\braket{v_{\ell}}{h_{1}} 
			\right\} \\
	& = & \bra{g_{2}} e^{itA_{G}} \ket{g_{1}} \bra{h_{2}} e^{-itA_{H}} \ket{h_{1}},
\end{eqnarray}
since $e^{-it(m-1)\lambda_{k}} = e^{it\lambda_{k}}$, for all $k$.
This shows that $G_{K_{m}}[H]$ has perfect state transfer from $(g_{1},h_{1})$ to $(g_{2},h_{2})$ at time $t$.
\qed\\

\par\noindent
The {\em standard} lexicographic product $G[H]$ is obtained when we let $C=J$ in Equation (\ref{eqn:generalized-lexico}).
In this case, Equation (\ref{eqn:pst-lexico}) decouples nicely and we have a similar result to 
Proposition \ref{prop:lexico-clique} but without requiring $G$ to have perfect state transfer.

\begin{lemma} \label{lemma:std-lexico}
Let $G$ be an arbitrary graph and
let $H$ be a regular graph with perfect state transfer at time $t_{H}$ from $h_{1}$ to $h_{2}$, for $h_{1},h_{2} \in V_{H}$.
Then, $G[H]$ has perfect state transfer from $(g,h_{1})$ to $(g,h_{2})$, for any $g \in V_{G}$, if
\begin{equation}
t_{H}|V_{H}|\Spec(G) \subseteq 2\ZZ\pi.
\end{equation}
\end{lemma}
\prf
If $H$ is an $m$-vertex regular graph, then $[A_{H},J_{m}] = 0$. 
The all-one matrix $J_{m}$ has eigenvalues $m$ (with multiplicity one) and $0$ (with multiplicity $m-1$).
Thus, Equation (\ref{eqn:pst-lexico}) becomes
\begin{eqnarray} \label{eqn:pst-lexico2}
\lefteqn{\bra{g,h_{2}}e^{-it_{H}A_{G[H]}}\ket{g,h_{1}}}\\
	& = & \sum_{k} \braket{g}{u_{k}}\braket{u_{k}}{g}
			\left\{ 
				e^{-it_{H}(\lambda_{k}m+\mu_{0})} \braket{h_{2}}{v_{0}}\braket{v_{0}}{h_{1}} 
				+
				\sum_{\ell \neq 0} e^{-it_{H}\mu_{\ell}} \braket{h_{2}}{v_{\ell}}\braket{v_{\ell}}{h_{1}} 
			\right\} \\
	& = & \bra{h_{2}} e^{-it_{H}A_{H}} \ket{h_{1}}, 
\end{eqnarray}
since $e^{-it_{H}m \lambda_{k}(G)}=1$, for all $k$, and $\sum_{k} \ket{u_{k}}\bra{u_{k}} = I$. This proves the claim.
\qed\\

\par\noindent{\em Remark}:
We will adopt the convention of scaling quantum walk time with respect to the {\em size} of the underlying graphs. 
Moore and Russell \cite{mr02} proved that a continuous-time quantum walk on the $n$-cube $Q_{n}$ has a uniform mixing time 
of $(2\ZZ + 1)\frac{\pi}{4}n$ (which shows the time scaling with respect to the dimension of the $n$-cube). 
They used $H = \frac{1}{n}A_{Q_{n}}$ as their Hamiltonian -- which is the probability transition matrix of 
the simple random walk on $Q_{n}$. 

\begin{corollary}
Suppose $H$ is a $k_{H}$-regular graph with perfect state transfer at time $t_{H} = \frac{\pi}{2}k_{H}$
and $G$ is an integral graph (all of its eigenvalues are integers). Then, $G[H]$ has perfect state transfer 
provided $k_{H}|V_{H}|\Spec(G) \subseteq 4\ZZ$.
\end{corollary}
\prf
Apply Lemma \ref{lemma:std-lexico} by noting that $e^{-it_{H}|V_{H}|\lambda_{k}(G)} = 1$, 
since $t_{H} = \frac{\pi}{2}k_{H}$ and $\lambda_{k}(G)k_{H}|V_{H}|$ is divisible by $4$, for all $k$.
\qed\\

\par\noindent
The $n$-cube $Q_{n}$ is a $n$-regular graph on $2^{n}$ vertices which has perfect state transfer at time 
$n\frac{\pi}{2}$ (with time scaling) (see \cite{bgs08}). Thus, for any integral graph $G$, the composition graph 
$G[Q_{n}]$ has perfect state transfer if $n \ge 2$.


\section{Cones}

In this section, we explore some constructions of perfect state transfer graphs which generalize the double cones
studied by Bose \etal \cite{bcms09} and Angeles-Canul \etal \cite{anoprt10,anoprt09}. The goal behind these constructions 
is to understand the types of intermediate graphs which allow perfect state transfer between the two antipodal vertices. 
For the double cones $\{K_{2},\overline{K}_{2}\}+\REG_{n,k}$, the intermediate graphs are $n$-vertex
$k$-regular graphs and sufficient conditions for perfect state transfer on $n$ and $k$ were derived 
in \cite{anoprt10}. 

Here, we consider more complex cones by allowing irregular graphs (on double cones), by increasing the number 
of intermediate layers, and by varying the connectivity structure (using semi-regular bipartite connections).
We show new perfect state transfer graphs for irregular double cones and for double half-cones with circulant connections,
and also prove negative results for longer diameter cones on join connections.

\subsection{Irregular double cones}

We recall the Perron-Frobenius theory of nonnegative matrices.
A matrix is called {\em nonnegative} if it has no negative entries.
The spectral radius of a matrix $A$, denoted $\rho(A)$, is the maximum eigenvalue of $A$ (in absolute value).
The Perron-Frobenius theorem for nonnegative matrices states that
if $A$ is a real nonnegative $n \times n$ matrix whose underlying directed graph $G$ is strongly connected,
then $\rho = \rho(A)$ is a simple eigenvalue of $A$; moreover, the unique eigenvector corresponding to $\rho$
has no zero entries and all entries have the same sign.

In what follows, we denote $K_{2}^{b}$ as the two-vertex graph 
which equals $K_{2}$ if $b=1$, and equals $\overline{K}_{2}$ if $b=0$.

\begin{theorem} \label{thm:double-cone}
Let $G$ be any connected graph whose maximum (simple) eigenvalue is $\lambda_{0}$ with a corresponding 
positive (normalized) eigenvector $\ket{x_{0}}$.
Consider the double cone $\GG = K_{2}^{b} + G$, for $b \in \{0,1\}$, where the edges adjacent to the vertices 
of $K_{2}^{b}$, say $A$ and $B$, are weighted proportional to $\alpha\ket{x_{0}}$.
Then, the fidelity between $A$ and $B$ is given by
\begin{equation}
\bra{B}e^{-itA_{\GG}}\ket{A} =
	\frac{1}{2}
	\left\{e^{-it\tilde{\lambda}_{0}^{+}}\left[\cos(t\Delta) + i\frac{\lambda_{0}^{-}}{\Delta}\sin(t\Delta)\right] - 1\right\},
\end{equation}
where $\tilde{\lambda}_{0}^{\pm} = (\lambda_{0} \pm b)/2$ and
$\Delta = \sqrt{(\tilde{\lambda}_{0}^{-})^{2} + 2\alpha^{2}}$.
Thus, perfect state transfer is achieved if
$\tilde{\lambda}_{0}^{+}/\Delta \in \QQ_{0,1} \cup \QQ_{1,0}$.
\end{theorem}
\prf
Let $A_{G}$ be the adjacency matrix of $G$. The adjacency matrix of $\GG$ is
\begin{equation}
A_{\GG} = 
\begin{bmatrix}
0 & b & \alpha\bra{x_{0}} \\
b & 0 & \alpha\bra{x_{0}} \\
\alpha\ket{x_{0}} & \alpha\ket{x_{0}} & A_{G}
\end{bmatrix}.
\end{equation}
For $1 \le k \le n-1$, let $\lambda_{k}$ and $\ket{x_{k}}$ be the other eigenvalues and eigenvectors of $A_{G}$.
Next, we define the following quantities:
\begin{equation}
\tilde{\lambda}_{0}^{\pm} = \frac{\lambda_{0} \pm b}{2},
\hspace{.2in}
\Delta = \sqrt{(\tilde{\lambda}_{0}^{-})^{2} + 2\alpha^{2}},
\hspace{.2in}
\lambda_{\pm} = \tilde{\lambda}_{0}^{+} \pm \Delta.
\hspace{.2in}
\kappa_{\pm} = \tilde{\lambda}_{0}^{-} \pm \Delta.
\end{equation}
The eigenvalues of $A_{\GG}$ are given by 
$\lambda_{0} = 0$, 
$\lambda_{\pm}$, 
and
$\lambda_{k}$, $1 \le k \le n-1$, 
with corresponding eigenvectors
\begin{equation}
\ket{z_{0}} = 
\frac{1}{\sqrt{2}}
\begin{bmatrix}
+1 \\ -1 \\ \ket{0_{n}}
\end{bmatrix},
\hspace{.1in}
\ket{z_{\pm}} =
\frac{1}{L_{\pm}}
\begin{bmatrix}
\alpha/\kappa_{\pm} \\ \alpha/\kappa_{\pm} \\ \ket{x_{0}}
\end{bmatrix},
\hspace{.1in}
\ket{z_{k}} =
\begin{bmatrix}
0 \\ 0 \\ \ket{x_{k}}
\end{bmatrix}
\end{equation}
where $L_{\pm}^{2} = 2\alpha^{2}/\kappa_{\pm}^{2} + 1$.
Note $(\kappa_{\pm} L_{\pm})^{2} = 2\alpha^{2} + \kappa_{\pm}^{2} = 2\Delta(\Delta \pm \tilde{\lambda}_{0}^{-})$.
The fidelity between $A$ and $B$, namely $\bra{B}e^{-itA_{\GG}}\ket{A}$, is given by
\begin{eqnarray}
\sum_{\pm} \frac{\alpha^{2}e^{-it\lambda_{\pm}}}{(\kappa_{\pm}L_{\pm})^{2}} - \frac{1}{2} 
	& = & \frac{1}{2}
		\left\{
		\frac{(\Delta-\tilde{\lambda}_{0}^{-})e^{-it(\tilde{\lambda}_{0}^{+} + \Delta)} + 
		(\Delta+\tilde{\lambda}_{0}^{-})e^{-it(\tilde{\lambda}_{0}^{+} - \Delta)}}{2\Delta}
		- 1 
		\right\} \\
	& = & \frac{1}{2}
		\left\{e^{-it\tilde{\lambda}_{0}^{+}}\left[ \cos(t\Delta) 
			+ i\frac{\tilde{\lambda}_{0}^{-}}{\Delta}\sin(t\Delta)\right] - 1 \right\}.
\end{eqnarray}
For perfect state transfer to occur, it is sufficient to have
$\tilde{\lambda}_{0}^{+}/\Delta \in \QQ_{0,1} \cup \QQ_{1,0}$.
\qed

\begin{figure}[t]
\begin{center}
\epsfig{file=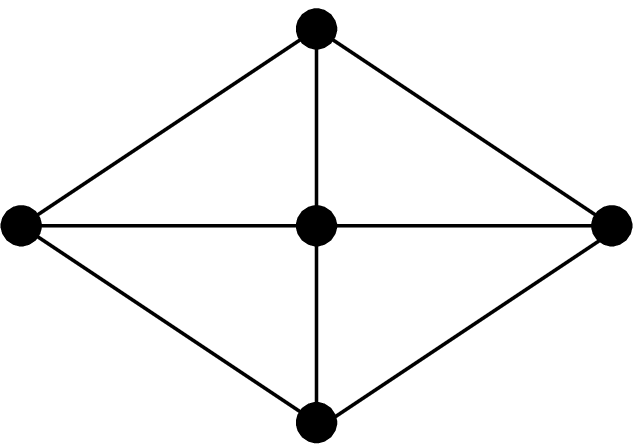, height=1.15in, width=1.5in}
\hspace{.5in}
\epsfig{file=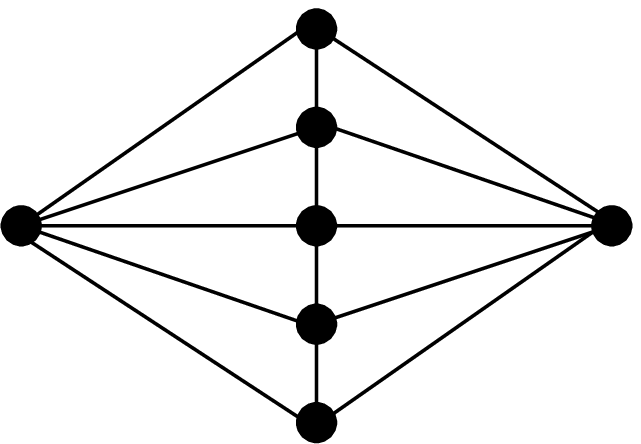, height=1.15in, width=1.5in}
\caption{Irregular weighted double cones have perfect state transfer: (a) $\overline{K}_{2} + P_{3}$;
(b) $\overline{K}_{2} + P_{5}$.}
\end{center}
\hrule
\end{figure}

\begin{corollary}
Let $G$ be any connected graph whose maximum (simple) eigenvalue is $\lambda_{0}$ 
with corresponding positive eigenvector $\ket{x_{0}}$.
Consider the double cone $\GG = \overline{K}_{2} + G$ where the edges adjacent to the two vertices of
$\overline{K}_{2}$, say $A$ and $B$, are weighted according to $\sqrt{n}\ket{x_{0}}$.
Then, perfect state transfer exists from $A$ to $B$ if
\begin{equation}
\frac{\lambda_{0}}{\sqrt{\lambda_{0}^{2} + 8n}} \ \in \ \QQ_{0,1} \cup \QQ_{1,0}.
\end{equation}
\end{corollary}
\prf 
In Theorem \ref{thm:double-cone} with $b=0$, let $\alpha = \sqrt{n}$ and note
$\tilde{\lambda}_{0}^{\pm} = \lambda_{0}/2$. Thus, 
$\tilde{\lambda}_{0}^{+}/\Delta = \lambda_{0}/\sqrt{\lambda_{0}^{2} + 8n}$, 
which proves the claim.
\qed\\

\par\noindent{\em Remark}:
Given $n$, we may choose $\lambda_{0} = \sqrt{8n/3}$ so the sufficient condition
$\lambda_{0}/\sqrt{\lambda_{0}^{2} + 8n} = 1/2$
is satisfied for perfect state transfer. Moreover, we can find a uniform edge weighting for $G$ so that 
$\sqrt{8n/3}$ is a dominant eigenvalue.
Thus, in the presence of weights, any double cone $\overline{K}_{2}+G$ has perfect state transfer.


\subsection{Glued double cones}

Analogous to the construction of glued-(binary)trees in Childs \etal \cite{ccdfgs03}, we consider 
gluing two double cones using a semi-regular bipartite connection to obtain a perfect state transfer graph. 
In contrast, gluing two double cones using the join (full bipartite) connection yields no perfect state transfer 
(even with weights) as proved in \cite{anoprt09}.

\begin{theorem} \label{thm:glued-cones}
Let $G \in \REG_{n,k}$ and let $C$ be a symmetric Boolean matrix which commutes with the adjacency matrix of $G$. 
Suppose that $C\ket{1_{n}} = \gamma\ket{1_{n}}$.
Let $k_{\pm} = \frac{1}{2}(k \pm \gamma)$ and $\Delta_{\pm} = \sqrt{k_{\pm} + n}$.
Then, the graph $\GG = K_{1}+G \circ G+K_{1}$, formed by taking two copies of $K_{1} + G$ and connecting the copies 
of $G$ using $C$, has perfect state transfer if
$\Delta_{+}/\Delta_{-} \in \QQ_{0,1} \cup \QQ_{1,0}$ and
at least one of $\gamma/\Delta_{+}$ or $\gamma/\Delta_{-}$ is in $\QQ_{0,1}$.
\end{theorem}
\prf
Suppose the eigenvalues and eigenvectors of $G$ are $\lambda_{k}$ and $\ket{v_{k}}$, respectively,
where $k = \lambda_{0} > \lambda_{1} \ge \ldots \ge \lambda_{n-1}$.
The adjacency matrix of $\GG$ is given by
\begin{equation}
A_{\GG} = 
\begin{bmatrix}
0 & 0 & \bra{1_{n}} & \bra{0_{n}} \\
0 & 0 & \bra{0_{n}} & \bra{1_{n}} \\
\ket{1_{n}} & \ket{0_{n}} & A_{G} & C \\
\ket{0_{n}} & \ket{1_{n}} & C & A_{G} 
\end{bmatrix}.
\end{equation}
Let $k_{\pm} = \frac{1}{2}(k \pm \gamma)$ and $\Delta_{\pm}^{2} = k_{\pm}^{2} + n$.
Let $\alpha_{\pm} = k_{+} \pm \Delta_{+}$
and $\beta_{\pm} = k_{-} \pm \Delta_{-}$.
The eigenvalues of $A_{\GG}$ are given by
$\alpha_{\pm}$, $\beta_{\pm}$, and $\pm\lambda_{k}$, for $k \neq 0$,
with corresponding eigenvectors: 
\begin{equation}
\ket{\alpha_{\pm}} = \frac{1}{L_{\pm}}\begin{bmatrix} 1 \\ 1 \\ \frac{1}{n}\alpha_{\pm}\ket{1_{n}} \\ \frac{1}{n}\alpha_{\pm}\ket{1_{n}} \end{bmatrix},
\hspace{.1in}
\ket{\beta_{\pm}} = \frac{1}{M_{\pm}}\begin{bmatrix} +1 \\ -1 \\ +\frac{1}{n}\beta_{\pm}\ket{1_{n}} \\ -\frac{1}{n}\beta_{\pm}\ket{1_{n}}\end{bmatrix},
\hspace{.1in}
\ket{\lambda_{k}} = \frac{1}{\sqrt{2}}\begin{bmatrix} 0 \\ 0 \\ \ket{v_{k}} \\ \pm\ket{v_{k}} \end{bmatrix},
\end{equation}
where $L_{\pm}^{2} = \frac{2}{n}(n + \alpha_{\pm}^{2})$ and
$M_{\pm}^{2} = \frac{2}{n}(n + \beta_{\pm}^{2})$ are the normalization constants.
The quantum walks between involving the cone vertices, say $A$ and $B$, are given by
\begin{eqnarray}
\bra{B}e^{-itA_{\GG}}\ket{A}
	& = & \sum_{\pm} \frac{e^{-it\alpha_{\pm}}}{L_{\pm}^{2}}
			- \sum_{\pm} \frac{e^{-it\beta_{\pm}}}{M_{\pm}^{2}} \\
\bra{A}e^{-itA_{\GG}}\ket{A}
	& = & \sum_{\pm} \frac{e^{-it\alpha_{\pm}}}{L_{\pm}^{2}}
			+ \sum_{\pm} \frac{e^{-it\beta_{\pm}}}{M_{\pm}^{2}}
\end{eqnarray}
At time $t=0$, the second equation yields
$1 = \sum_{\pm} {L_{\pm}^{-2}} + \sum_{\pm} {M_{\pm}^{-2}}$.
To achieve perfect state transfer, it suffices to require
\begin{equation}
e^{-it\alpha_{\pm}} =  +1,
\hspace{.1in}
e^{-it\beta_{\pm}} = -1,
\hspace{.1in}
e^{-it\gamma/2} = \pm 1.
\end{equation}
We may restate these conditions as
$\Delta_{+}/\Delta_{-} = \QQ_{0,1} \cup \QQ_{1,0}$
and $\{\gamma/\Delta_{+},\gamma/\Delta_{-}\} \cap \QQ_{0,1} \neq \emptyset$.
\qed\\

\begin{figure}[t]
\begin{center}
\epsfig{file=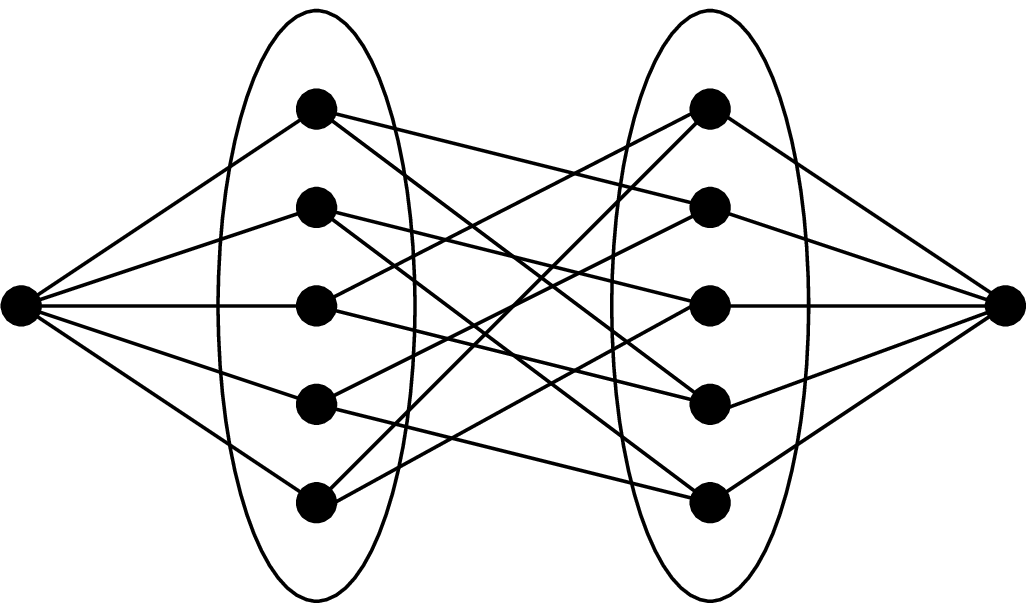, height=1.25in, width=2in}
\hspace{.5in}
\epsfig{file=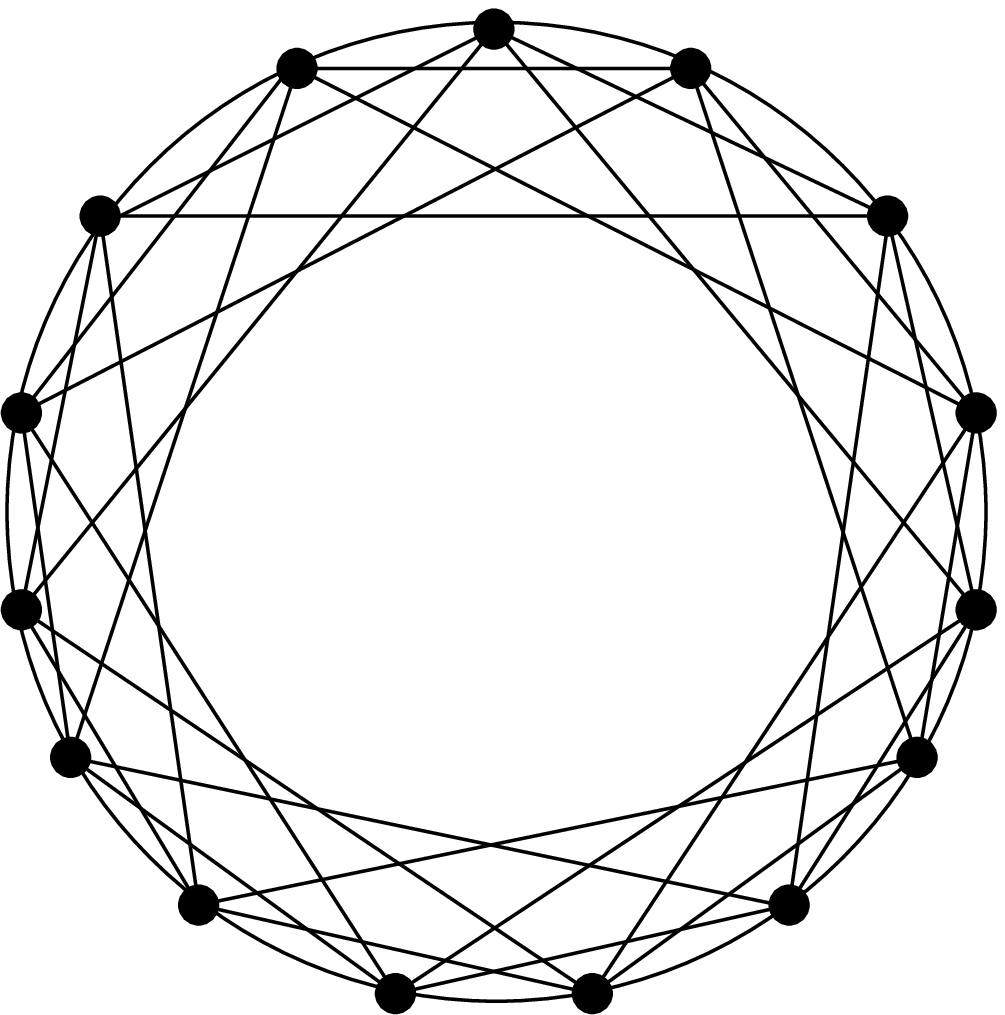, height=1.5in, width=1.5in}
\hspace{.5in}
\epsfig{file=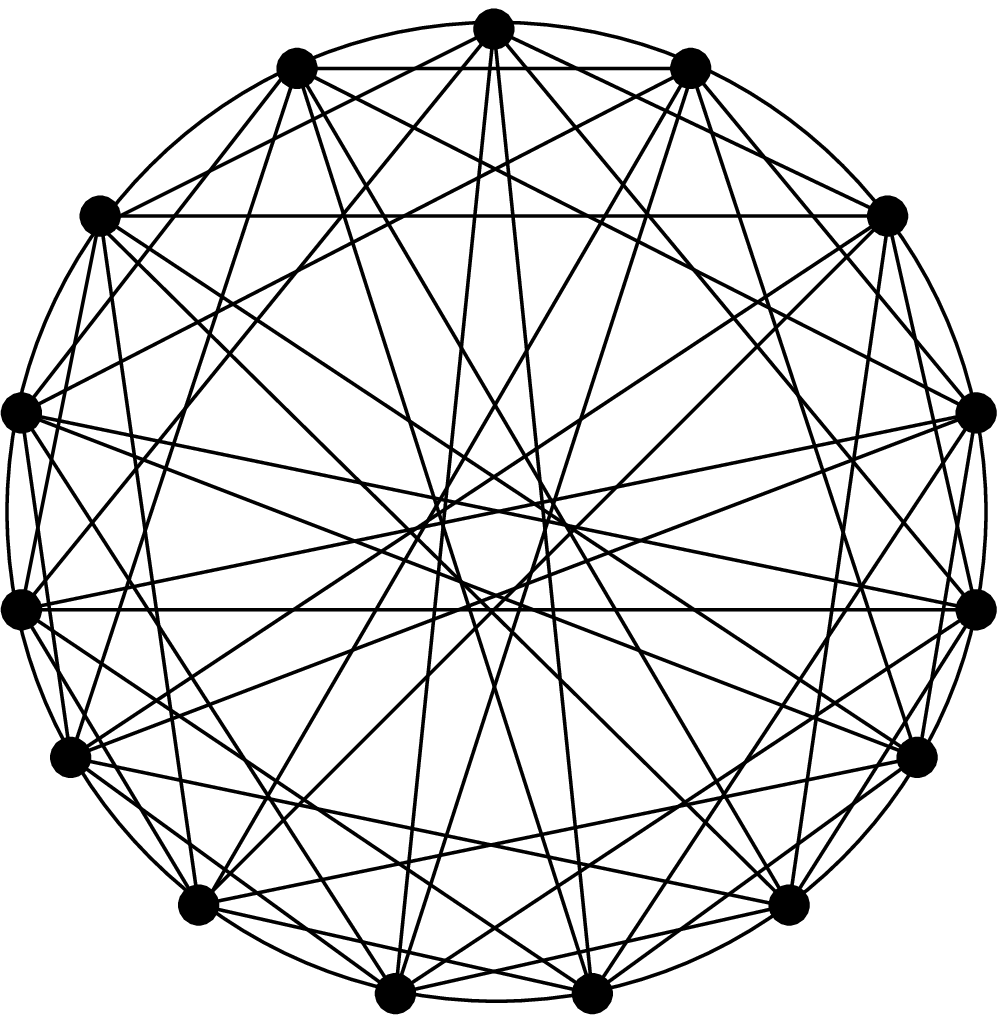, height=1.5in, width=1.5in}
\caption{Glued cones (a) $K_{1}+G \circ G+K_{1}$ has perfect state transfer, with 
(b) $G=Circ(15,\{1,2,4\})$; (c) $C=Circ(15,\{1,2,4,7\})$. The connection $\circ$ is defined by $C$.
}
\label{fig:P4-join}
\end{center}
\hrule
\end{figure}

\par\noindent{\em Remark}:
In Theorem \ref{thm:glued-cones}, the result also holds if we replace $G$ with two distinct graphs 
$G_{1}$ and $G_{2}$ from the same family $\REG_{n.k}$.\\

In the following corollary, we describe an explicit family of glued double cones which exhibit
perfect state transfer. The construction uses a pair of circulant families of graphs (see Figure \ref{fig:P4-join}).

\begin{corollary}
For $a \ge 2$, let $n = 15 \times 2^{2(a-2)}$, $k = 3 \times 2^{a-1}$, and $\gamma = 4 \times 2^{a-1}$.
Consider two circulant graphs $G = Circ(n,[k/2])$ and $C = Circ(n,[\gamma/2])$.
Then, the graph $\GG = K_{1}+G \circ G+K_{1}$ has perfect state transfer, where the connection $\circ$
is specified by $C$.
\end{corollary}
\prf
Note we have
$k_{\pm} = \frac{1}{2}(k \pm \gamma) = 2^{a-2}(3 \pm 4)$ and
$\Delta_{\pm} = 2^{a-2}((3 \pm 4)^{2} + 15) \in 2^{a-2}\{8,4\}$.
Thus, $\Delta_{+}/\Delta_{-} = 2 \in \QQ_{0,1}$ and
$\gamma/\Delta_{-} = 2 \in \QQ_{0,1}$, which satisfy the sufficiency conditions for
perfect state transfer in Theorem \ref{thm:glued-cones}.
\qed


\subsection{Cylindrical cones}

In this section, we consider graphs of the form $K_{1} + G_{1} + H + G_{2} + K_{1}$, where $G_{1},G_{2} \in \REG_{n,k}$
and $H \in \REG_{m,\ell}$. We show a negative result for perfect state transfer whenever $H$ is the empty graph.
This generalizes known negative results on $P_{4}$ and $K_{1}+G+G+K_{1}$ (see \cite{cddekl05,anoprt09}).

\begin{theorem} \label{thm:p5-join-pst}
For any integers $n,k,m$ where $n \ge 1$, $0 \le k < n$, and $m \ge 1$, 
the graph $K_{1} + G_{1} + \overline{K}_{m} + G_{2} + K_{1}$ has no perfect state transfer,
whenever $G_{1},G_{2} \in \REG_{n,k}$.
\end{theorem}
\prf
Let $\GG$ be the graph $K_{1} + G_{1} + H + G_{2} + K_{1}$, where $G_{1},G_{2} \in \REG_{n,k}$ and $H \in \REG_{m,\ell}$.
Let $A_{G_{1}}$ be the adjacency matrix of $G_{1}$ with eigenvalues $\alpha_{r}$ and eigenvectors $\ket{u_{r}}$; similarly,
let $A_{G_{2}}$ be the adjacency matrix of $G_{2}$ with eigenvalues $\beta_{r}$ and eigenvectors $\ket{v_{r}}$,
for $r \in [n]$. Note $k = \alpha_{0} = \beta_{0}$ are the simple maximum eigenvalues of both $G_{1}$ and $G_{2}$.
Let $A_{H}$ be the adjacency matrix of $H$ with eigenvalues $\rho_{s}$ and eigenvectors $\ket{w_{s}}$,
where $\ell = \rho_{0}$ is the simple maximum eigenvalue of $H$.
Thus, the adjacency matrix of $\GG$ is given by
\begin{equation}
A_{\GG} = 
\begin{bmatrix}
0 & 0 & \bra{1_{n}} & \bra{0_{n}} & \bra{0_{n}} \\
0 & 0 & \bra{0_{n}} & \bra{0_{n}} & \bra{1_{n}} \\
\ket{1_{n}} & \ket{0_{n}} & A_{G_{1}} & J_{n,m} & O_{n,n} \\
\ket{0_{m}} & \ket{0_{m}} & J_{m,n} & A_{H} & J_{m,n} \\
\ket{0_{n}} & \ket{1_{n}} & O_{n,n} & J_{n,m} & A_{G_{2}} 
\end{bmatrix}.
\end{equation}
In our case, we have $A_{H} = O_{m,m}$ is the zero $m \times m$ matrix and $\ell = 0$.

Let $\lambda_{\pm}$ be the roots of quadratic polynomial $\lambda^{2} - k\lambda - n = 0$;
thus $\lambda_{\pm} = \tilde{k} \pm \Delta$, 
where $\tilde{k} = k/2$ and $\Delta^{2} = \tilde{k}^{2} + n$.
Consider roots of the cubic polynomial $(\mu - \ell)(\mu^{2} - k\mu - (2m+1)n) - 2\ell mn = 0$.
For $\ell = 0$, zero is a root of this cubic along with the two roots of the quadratic equation 
$\mu^{2} - k\mu - (2m+1)n = 0$. Let $\mu_{\pm} = \tilde{k} \pm \Gamma$, 
where $\Gamma^{2} = \tilde{k}^{2} + (2m+1)n$.
The eigenvalues of $A_{\GG}$ are given by
$\lambda_{\pm}$, $\mu_{\pm}$, $0$, and $\lambda^{(1)}_{r}$, $\lambda^{(3)}_{r}$, for $r \neq 0$, 
and $\lambda^{(2)}_{s}$, for $s \neq 0$, with corresponding eigenvectors: 
\begin{equation}
\ket{\lambda_{\pm}} 
	= \frac{1}{L_{\pm}}
	\begin{bmatrix} +1 \\ -1 \\ +\frac{1}{n}\lambda_{\pm}\ket{1_{n}} \\ \ket{0_{n}} \\ -\frac{1}{n}\lambda_{\pm}\ket{1_{n}} \end{bmatrix},
\hspace{.1in}
\ket{\mu_{\pm}} 
	= \frac{1}{M_{\pm}}
	\begin{bmatrix} 1 \\ 1 \\ \frac{1}{n}\mu_{\pm}\ket{1_{n}} \\ 2\ket{1_{n}} \\ \frac{1}{n}\mu_{\pm}\ket{1_{n}} \end{bmatrix},
\hspace{.1in}
\ket{v_{0}}
	= \frac{1}{N}
	\begin{bmatrix} 1 \\ 1 \\ \ket{0_{n}} \\ -1/m\ket{1_{n}} \\ \ket{0_{n}} \end{bmatrix},
\end{equation}
and
\begin{equation}
\ket{\lambda^{(1)}_{r}} = \begin{bmatrix} 0 \\ 0 \\ \ket{u_{r}} \\ \ket{0_{m}} \\ \ket{0_{n}} \end{bmatrix},
\hspace{.1in}
\ket{\lambda^{(2)}_{s}} = \begin{bmatrix} 0 \\ 0 \\ \ket{0_{n}} \\ \ket{w_{s}} \\ \ket{0_{n}} \end{bmatrix},
\hspace{.1in}
\ket{\lambda^{(3)}_{r}} = \begin{bmatrix} 0 \\ 0 \\ \ket{0_{n}} \\ \ket{0_{m}} \\ \ket{v_{r}} \end{bmatrix},
\end{equation}
where $1 \le r < n$ and $1 \le s < m$. 
Here $L_{\pm}$, $M_{\pm}$ and $N$ are normalization factors.
We have the following fidelities:
\begin{eqnarray} 
\label{eqn:A2B}
\bra{B}e^{-itA_{\GG}}\ket{A}
	& = & -\sum_{\pm} \frac{e^{-it\lambda_{\pm}}}{L_{\pm}^{2}}
			+ \sum_{\pm} \frac{e^{-it\mu_{\pm}}}{M_{\pm}^{2}} 
			+ \frac{1}{N^{2}} \\
\label{eqn:A2A}
\bra{A}e^{-itA_{\GG}}\ket{A}
	& = & \sum_{\pm} \frac{e^{-it\lambda_{\pm}}}{L_{\pm}^{2}}
			+ \sum_{\pm} \frac{e^{-it\mu_{\pm}}}{M_{\pm}^{2}} 
			+ \frac{1}{N^{2}}.
\end{eqnarray}
At time $t=0$, Equation (\ref{eqn:A2A}) yields
\begin{equation}
1 = \sum_{\pm} \frac{1}{L_{\pm}^{2}} + \sum_{\pm} \frac{1}{M_{\pm}^{2}} + \frac{1}{N^{2}}.
\end{equation}
So, to achieve perfect state transfer in Equation (\ref{eqn:A2B}), we require that
\begin{equation}
e^{-it\lambda_{\pm}} = -1,
\hspace{.1in}
e^{-it\mu_{\pm}} = +1.
\end{equation}
This implies $t(\tilde{k} \pm \Delta) \in (2\ZZ+1)\pi$ and $t(\tilde{k} \pm \Gamma) \in (2\ZZ)\pi$.
We restate these conditions as
\begin{equation}
\frac{\tilde{k} \pm \Delta}{\tilde{k} \pm \Gamma} \in \mathbb{Q}_{1,0},
\hspace{.2in}
\frac{\tilde{k} \pm \Delta}{\tilde{k} \mp \Gamma} \in \mathbb{Q}_{1,0}.
\end{equation}
Clearly it is necessary to have $\Delta,\Gamma\in \mathbb{Z}$, else the above quotients are not even rational.

Observe that if both $\tilde{k}$ and $n$ are odd, both quotients lie in $\mathbb{Q}_{1,1}$. 
If $\tilde{k}$ is even and $n$ is odd, the same is true. 
If $\tilde{k}$ is odd and $n$ is even, then the numerator and denominator of at least one of the quotients 
must be congruent to 2 modulo 4, and so one lies in $\mathbb{Q}_{1,1}$. 
If both $\tilde{k}$ and $n$ are even, than we can divide the numerator and denominator of each quotient 
by 2 (clearly, then, 4 divides $n$ as well), rewriting the conditions as:
\begin{equation}
\frac{\tilde{k}' \pm \Delta'}{\tilde{k}' \pm \Gamma'} \in \mathbb{Q}_{1,0},
\hspace{.2in}
\frac{\tilde{k}' \pm \Delta'}{\tilde{k}' \mp \Gamma'} \in \mathbb{Q}_{1,0},
\end{equation}
where $\tilde{k}' = \tilde{k}/2$, $\Delta' = \Delta/2$, and $\Gamma' = \Gamma/2$. 
Since this is in essence the same set of conditions as before, we argue by infinite descent 
that there can be no solutions of this form. Since we have ruled out all parity combinations 
for $\tilde{k}$ and $n$, there can be no solutions and no perfect state transfer in this case.
\qed

\begin{figure}[t]
\begin{center}
\epsfig{file=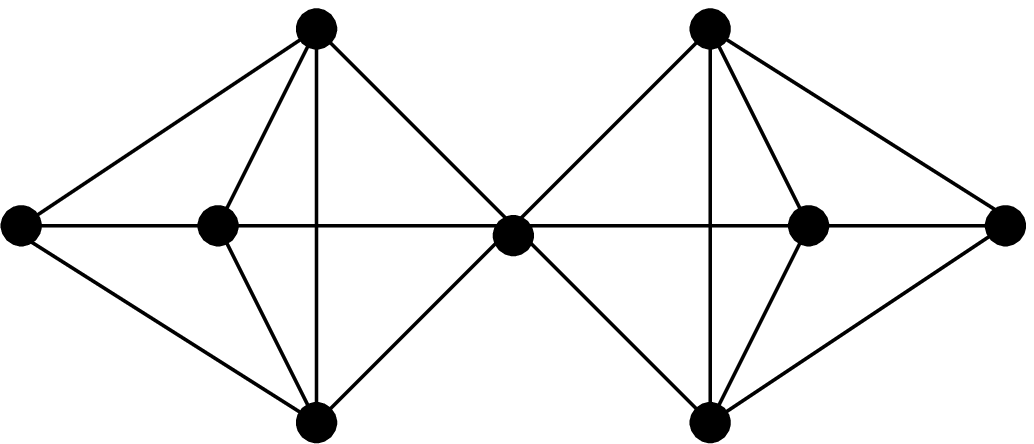, height=1in, width=2.05in}
\hspace{.2in}
\epsfig{file=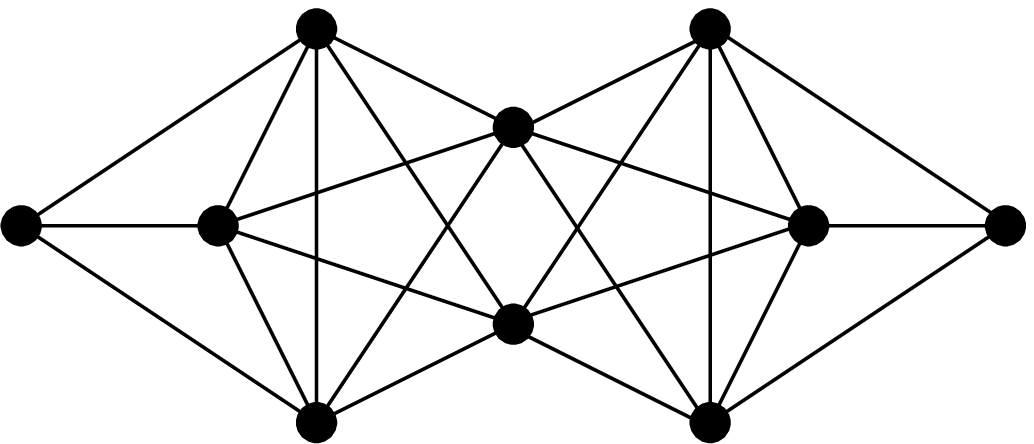, height=1in, width=2.05in}
\caption{Cylindrical cones of diameter five with no perfect state transfer: 
(a) $K_{1} + K_{3} + K_{1} + K_{3} + K_{1}$; 
(b) $K_{1} + K_{3} + \overline{K}_{2} + K_{3} + K_{1}$.}
\label{fig:P5-join}
\end{center}
\hrule
\end{figure}


\section{Equitable partitions}

The path-collapsing argument was used by Christandl \etal \cite{cddekl05} to show that weighted paths 
have perfect state transfer. 
This follows because the (unweighted) $n$-dimensional hypercube $Q_{n}$ has perfect state transfer and 
it can be collapsed to a weighted path. On the other hand, this argument was used in the opposite direction
by Childs \etal \cite{ccdfgs03} to show that a continuous-time quantum walk on an unweighted layered graph 
has polynomial hitting time by observing its behavior on a corresponding weighted path.

A natural way to view this reduction is by using equitable distance partitions and their quotient graphs 
(for example, see \cite{godsil93,kb08}). But, most quotient graphs derived this way are directed and hence not 
suitable for quantum walks. The path-collapsing reduction offers a way to {\em symmetrize} these directed quotient 
graphs into undirected graphs. In what follows, we formalize and generalize this argument using the theory of 
equitable partitions (see Godsil \cite{godsil93}).

\begin{lemma} \label{lemma:generalized-path-collapsing}
Let $G=(V,E)$ be a graph with an equitable distance partition $\pi = \biguplus_{j=0}^{m-1} V_{j}$ 
with respect to vertices $a$ and $b$.
Then, the fidelity of a quantum walk on $G$ between vertices $a$ and $b$
is equivalent to the fidelity of a quantum walk on a symmetrized quotient graph $G/\pi$ between 
$\pi(a)=V_{0}$ and $\pi(b)=V_{m-1}$;
namely, if $B_{G/\pi}[j,k] = \sqrt{d_{j,k}d_{k,j}}$, for all $j,k \in [m]$, then
\begin{equation}
|\bra{b}e^{-itA_{G}}\ket{a}| = |\bra{\pi(b)}e^{-itB_{G/\pi}}\ket{\pi(a)}|.
\end{equation}
\end{lemma}
\prf
For $j,k \in [m]$, let $d_{j,k}$ be the number of vertices in $V_{k}$ adjacent to each vertex $x$ in $V_{j}$.
Let $P$ be the characteristic partition $n \times m$ matrix of $\pi$; namely,
$P[j,\ell] = 1$ if vertex $j$ belongs to partition $V_{\ell}$, and $0$ otherwise.
Suppose $Q$ be the matrix $P$ after we normalize each column; so $Q^{T}Q = I_{n}$.
Then, we have
\begin{equation}
A_{G} Q = Q B_{G/\pi},
\end{equation}
where
\begin{equation}
B_{G/\pi}[j,k] = \sqrt{d_{j,k}d_{k,j}}.
\end{equation}
The matrix $B_{G/\pi}$ is defined implicitly in \cite{cddekl05} through the columns of $Q$
(viewed as basis states in a new graph)\footnote{Note $B_{G/\pi}$ is different from $A_{G/\pi}$ 
(as defined in \cite{godsil93}), since $B_{G/\pi}$ is symmetric and represents an undirected weighted graph
whereas $A_{G/\pi}$ represents a directed graph.}.
The following spectral correspondences between $A_{G}$ and $B_{G/\pi}$ can be shown:
\begin{itemize}
\item If $A_{G}\ket{y} = \lambda\ket{y}$, then $B_{G/\pi}\ket{x} = \lambda\ket{x}$, where $\ket{x} = Q^{T}\ket{y}$,
	provided $Q^{T}\ket{y} \neq 0$.
\item If $B_{G/\pi}\ket{x} = \lambda\ket{x}$, then $A_{G}\ket{y} = \lambda\ket{y}$, where $\ket{y} = Q\ket{x}$.
\end{itemize}
Suppose that $E(A_{G}) = \{\ket{y_{k}} : k \in [n]\}$ is the (orthonormal) set of eigenvectors of $A_{G}$;
similarly, let $E(B_{G/\pi}) = \{\ket{x_{k}} : k \in [m]\}$ be the (orthonormal) set of eigenvectors of $B_{G/\pi}$.
Since $\pi(a) = \{a\}$ and $\pi(b) = \{b\}$ are singleton partitions, 
we have	$Q^{T}\ket{a} = \ket{\pi(a)}$ and $Q^{T}\ket{b} = \ket{\pi(b)}$.
Thus, we have
\begin{eqnarray}
\bra{\pi(b)}e^{-itB_{G/\pi}}\ket{\pi(a)}
	& = & \bra{\pi(b)} \sum_{k=0}^{m-1} \left( e^{-it\lambda_{k}} \ket{x_{k}}\bra{x_{k}} \right) \ket{\pi(a)} \\
	& = & \bra{b} \sum_{k=0}^{m-1} \left( e^{-it\lambda_{k}} Q\ket{x_{k}}\bra{x_{k}}Q^{T} \right) \ket{a} \\
	& = & \bra{b}e^{-itA_{G}}\ket{a}.
\end{eqnarray}
The last step holds since the orthonormal eigenvectors of $A_{G}$ can be divided into two types: 
those that are constant on cells of $\pi$ (the ones of the form $\ket{y_{k}} = Q\ket{x_{k}}$, for some
eigenvector $\ket{x_{k}}$ of $B_{G/\pi}$) and those that sum to zero on each cell of $\pi$. The eigenvectors
of the latter type do not contribute to the quantum walk between the antipodal vertices $a$ and $b$.
\qed\\

\par\noindent{\em Remark}:
Lemma \ref{lemma:generalized-path-collapsing} shows that the double cones $\overline{K}_{2}+G$, for regular graphs 
$G \in \REG_{n,k}$, which have diameter two, are {\em equivalent} (in the sense of the fidelity of quantum walks between 
the antipodal vertices) to a weighted $P_{3}$ with adjacency matrix $\tilde{A}_{1}$ (shown below).
\begin{equation}
\tilde{A}_{1} =
\begin{bmatrix}
0 & \sqrt{n} & 0 \\
\sqrt{n} & k & \sqrt{n} \\
0 & \sqrt{n} & 0
\end{bmatrix}
\hspace{.2in}
\tilde{A}_{2} =
\begin{bmatrix}
0 & \sqrt{n} & 1 \\
\sqrt{n} & k & \sqrt{n} \\
1 & \sqrt{n} & 0
\end{bmatrix}
\end{equation}
The case of the connected double cone $K_{2}+G$, where $G \in \REG_{n,k}$, can also be shown to be equivalent to the
weighted graph with adjacency matrix $\tilde{A}_{2}$ (shown above).
This simplifies the analyses on values of $n$ and $k$ which allows perfect state transfer (see \cite{anoprt10}).\\

In what follows, we use the generalized path-collapsing argument above to revisit (unweighted) graphs of diameter three 
and compare them to (weighted) paths of length four. Then, we compare a family of symmetrically weighted paths $P_{4}$ 
(without self-loops) with a construction based on weak products.
This symmetry restriction on the weights can be made without loss of generality; see Kay \cite{kay09}.

\begin{lemma} \label{lemma:p4-gamma-kappa}
Let $P_{4}(\gamma;\kappa)$ denote a weighted path whose middle edge has weight $\gamma$ while the other two edges have unit weights
and whose two internal vertices have self-loops with weight $\kappa$ each.
Let $\Delta_{\pm} = \frac{1}{2}\sqrt{(\kappa \pm \gamma)^{2} + 4}$.
Then, $P_{4}(\gamma;\kappa)$ has perfect state transfer if:
\begin{enumerate}
\item Case $\kappa \neq 0$: $\Delta_{+}/\Delta_{-} \in \QQ_{0,1} \cup \QQ_{1,0}$ and
	$\{\gamma/\Delta_{+}, \gamma/\Delta_{-}\} \cap (\QQ_{0,1} \cup \QQ_{1,1}) \neq \emptyset$; or
\item Case $\kappa = 0$: $\{\gamma/\Delta_{+}, \gamma/\Delta_{-}\} \subseteq \QQ_{1,1}$ or
	$\{\gamma/\Delta_{+}, \gamma/\Delta_{-}\} \subseteq \QQ_{1,0}$.
\end{enumerate}
\end{lemma}
\prf
Let $k_{\pm} = (\kappa \pm \gamma)/2$, 
$\Delta_{+}^{2} = k_{+}^{2} + 1$,
and
$\Delta_{-}^{2} = k_{-}^{2} + 1$.
The adjacency matrix $A$ of $P_{4}(\gamma;\kappa)$, 
whose eigenvalues are $\alpha_{\pm} = k_{+} \pm \Delta_{+}$
and $\beta_{\pm} = k_{-} \pm \Delta_{-}$,
and its corresponding eigenvectors $\ket{\alpha_{\pm}}$ and $\ket{\beta_{\pm}}$ are given by:
\begin{equation}
A = 
\begin{bmatrix}
0  &  1      &  0      &  0 \\
1  &  \kappa &  \gamma &  0 \\
0  &  \gamma &  \kappa &  1 \\
0  &  0      &  1      &  0
\end{bmatrix},
\hspace{.25in}
\ket{\alpha_{\pm}} = 
\frac{1}{L_{\pm}}
\begin{bmatrix}
1 \\
1 \\
\alpha_{\pm} \\
\alpha_{\pm}
\end{bmatrix},
\hspace{.25in}
\ket{\beta_{\pm}} = 
\frac{1}{M_{\pm}}
\begin{bmatrix}
+1 \\
-1 \\
+\beta_{\pm} \\
-\beta_{\pm}
\end{bmatrix},
\end{equation}
where $L_{\pm}^{2} = 4\Delta_{+}(\Delta_{+} \pm k_{+})$ and
$M_{\pm}^{2} = 4\Delta_{-}(\Delta_{-} \pm k_{-})$.
The end-to-end fidelity of the quantum walk on $P_{4}(\gamma;\kappa)$ is given by
\begin{equation}
\bra{3}e^{-itP_{4}(\gamma;\kappa)}\ket{0} =
	\sum_{\pm} \frac{e^{-it\alpha_{\pm}}}{L_{\pm}^{2}} 
	- 
	\sum_{\pm} \frac{e^{-it\beta_{\pm}}}{M_{\pm}^{2}} 
\end{equation}
At time $t=0$, $\bra{0}e^{-itP_{4}(\gamma;\kappa)}\ket{0}$ equals
$\sum_{\pm} 1/L_{\pm}^{2} + \sum_{\pm} 1/M_{\pm}^{2} = 1$.
Thus, to achieve unit fidelity when $\kappa \neq 0$, it suffices to have $e^{-it\gamma/2}=\pm 1$, $\cos(t\Delta_{+})=\pm 1$, 
and $\cos(t\Delta_{-})=\mp 1$ where $t\Delta_{+}$ and $t\Delta_{-}$ differ in their parities (as a multiple of $\pi$)
while $t\gamma$ is of even parity. But, when $\kappa = 0$, and thus $\Delta_{+} = \Delta_{-}$, it suffices to simply have
$t\gamma$ be of odd parity.
\qed\\

\par\noindent
Note Theorem \ref{thm:glued-cones} forms a special case of Lemma \ref{lemma:p4-gamma-kappa} when $\kappa \neq 0$.
The fact that the analyses are equivalent follows from Lemma \ref{lemma:generalized-path-collapsing}.\\

\par\noindent{\em Remark}:
Let $P_{4}(\gamma)$ denote $P_{4}(\gamma;0)$; that is, a weighted path with no self-loops.
In this case, $\Delta_{+} = \Delta_{-}$ and a sufficient perfect state transfer condition is
$\Delta_{+}/\gamma \in \QQ_{1,1} \cup \QQ_{0,1}$.
\ignore{
For the second type, we require $\gamma/\Delta = p/q$, for an odd integer $p = 2k+1$
and an even integer $q = 2\ell$. More specifically, if $\ts$ is the perfect state transfer time, 
then $\gamma\ts = (2k+1)\pi$ and $\Delta\ts = 2\ell\pi$. Using the common time $\ts$, we get
\begin{equation}
\ts = \frac{(2k+1)\pi}{\gamma} = \frac{2\ell\pi}{\Delta}.
\end{equation}
Using the definition of $\Delta$, we get
$\gamma^{2} = [(2k+1)^{2}/4\ell^{2}](1 + \gamma^{2}/4)$.
}
So, $P_{4}(\gamma)$ has end-to-end perfect state transfer if either:
\begin{itemize}
\item for odd integer $K$ and even integer $L$, with $L > K$, we have $\gamma = 2\sqrt{K^{2}/(L^{2} - K^{2})}$; or
\item for odd integers $K$ and $L$, with $2L > K$, we have $\gamma = 2\sqrt{K^{2}/(4L^{2} - K^{2})}$.
\end{itemize}
The weak product $K_{2} \times K_{4k}$, for $k \ge 1$, has perfect state transfer if $m$ is divisible by $4$, 
by Proposition \ref{prop:weak-pst} (see Figure \ref{fig:graph-product}(a)). 
The path collapsing argument shows $K_{2} \times K_{4k}$ is equivalent to $P_{4}(\tau)$ where $\tau = (4k-2)/\sqrt{4k-1} > 1$. 
Thus, for perfect state transfer, the weak product construction $K_{2} \times K_{4k}$ yields edge weights greater than $1$, 
whereas $P_{4}(\gamma)$ can yield edge weights smaller than $1$ (with longer PST times).
We are not aware of unweighted constructions which can emulate the latter property.


\section{Conclusions}

Using the Cartesian graph product, Christandl \etal \cite{cddekl05} constructed two families of perfect state 
transfer graphs with large diameter, namely, $Q_{n}$ and $P_{3}^{\oplus n}$.
They also showed that weighted paths have perfect state transfer by a path-collapsing reduction from $Q_{n}$. 
This argument was used specifically on graphs with equitable distance partitions whose cells are empty graphs.
Our original motivation was to generalize the Cartesian product construction and extend the path-collapsing
argument to larger classes of graphs.

In this work, we described new families of graphs with perfect state transfer using the weak graph product 
and a generalized lexicographic product (which includes the Cartesian graph product as a special case). 
We also considered constructions involving double cones which allow the cell partitions to be non-empty graphs
(unlike the Cartesian product graphs). 
Here, we prove perfect state transfer on double cones of irregular graphs and on double half-cones connected 
by circulants. These generalized results in \cite{bcms09,anoprt10} on double cones of regular graphs and 
complement the negative result on double half-cones in \cite{anoprt09}. Although these cone constructions 
involve small diameter graphs, they provided insights into which intermediate graphs allow antipodal 
perfect state transfer. Non-antipodal perfect state transfer can also be derived from certain cones
(as shown in \cite{anoprt10}).

We also generalized the path-collapsing argument using the theory of equitable partitions. This can be used to 
show that certain weighted paths with self-loops have perfect state transfer. A possible interesting direction is 
to study random graphs with equitable distance partitions (as in the Anderson model \cite{anderson58}). 
A {\em weighted} path-collapsing argument would also be interesting since it can be used to analyze graphs produced 
in Feder's intriguing construction \cite{f06}.
The most elusive graph not covered by this framework is $P_{3}^{\oplus n}$ since none of the path-collapsing arguments
apply. This is because the connections are irregular (see Figure \ref{fig:cartesian}(b)). 
We leave these as open questions.


\section*{Acknowledgments}

The research was supported in part by the National Science Foundation grant DMS-1004531
and also by the National Security Agency grant H98230-09-1-0098.
We thank Richard Cleve, David Feder, and Michael Underwood for their helpful comments on perfect state transfer.



\end{document}